\documentclass[smallabstract,smallcaptions]{dccpaper}
\usepackage[utf8]{inputenc}
\usepackage{epsfig}
\usepackage{amsmath}
\usepackage{amssymb}
\usepackage{color}
\usepackage{url}
\usepackage{makeidx}  
\usepackage[ruled,linesnumbered]{algorithm2e}
\usepackage{url,enumitem}
\usepackage{multirow}
\usepackage{float}
\usepackage{url,enumitem}

\newcommand{\lra}{\leftarrow}
\newtheorem{defin}{Definition}
\newtheorem{example}{Example}

\title{Efficient Integer Retrieving from Unordered Compressed Sequences}

\begin{document}

\author{%
Igor~O.~Zavadskyi\\[0.5em]
{\small\begin{minipage}{\linewidth}\begin{center}
\begin{tabular}{c}
Taras Shevchenko National University of Kyiv \\
4d Glushkov Ave., Kyiv, Ukraine \\
\url{ihorzavadskyi@knu.ua}
\end{tabular}
\end{center}\end{minipage}}
}

\maketitle

\begin{abstract}
The variable-length Reverse Multi-Delimiter (RMD) codes are known to represent sequences of unbounded and unordered integers. When applied to data compression, they combine a good compression ratio with fast decoding. In this paper, we investigate another property of RMD-codes - the ability of direct access to codewords in the encoded bitstream. We present the method allowing us to extract and decode a codeword from an RMD-bitstream in almost constant time with the tiny space overhead, and make experiments on its application to natural language text compression.
\end{abstract}

\section{Introduction}

A compressed representation of integer sequences is the key element of different data compression techniques. It is used in frequency-based compression, when alphabet symbols are numbered so that smaller numbers are assigned to more frequent symbols, in the compact representation of suffix trees and arrays, offsets and lengths in LZ77-type compression, just to mention a few areas. In most cases the distribution of integers is biased in the direction of smaller ones. The variable length codes (VLC) provide a simple and space-efficient solution of the given problem. However, often not only compression itself is required but also performing different operations on compressed integer sequence, such as sequential decoding or extracting the element with a given index. While VLC are well-suited for sequential decoding, the direct access to elements of a VLC-bitstream is not obvious and straightforward and requires using auxiliary data structures and/or a special code construction.

If integers are arranged in ascending order and deltas between them are small enough, the problem of a direct access is reduced to performing the \textit{select} operation on a bitmap, where $i$-th bit is set if the number $i$ belongs to the sequence (as usual, we denote by $select(B,i)$ the position of the $i$-th one, while $rank(B,i)$ is equal to the number of ones from the beginning of the bitstream $B$ up to position $i$). There are a lot of solutions concerning rank and select on bitmaps; the overview of recent results can be found in \cite{kulekci} and \cite{pibiri}. Less attention was paid to extracting an element of an unordered number sequence given in a compressed form, although this operation is quite useful, e.g. in manipulating compressed texts or calculating values of $\Psi$ function used in compressed suffix arrays. Construction data structures for space-efficient representation of unordered integer sequences allowing the fast access to their elements is the goal of this paper.

Generally, we can distinguish 5 approaches to solve the mentioned problem more or less efficiently in terms of space and time. Here and below, we denote by $n$ the length of the integer sequence and assume the RAM memory model is used allowing constant time reading values from memory.

\begin{enumerate}[itemindent=1cm,leftmargin=0cm,itemsep=0cm,parsep=0.1cm]
    \item Encode the sequence using universal codes, such as Elias codes \cite{Elias}. Sample each $h$-th codeword and store pointers to sampled elements. To get the $x$-th element, obtain the position of the $\lfloor x/h\rfloor$-th sampled element and then search the bitstream sequentially. This approach is quite straightforward and requires $O(h)$ time overhead to access an element.
    \item Encode numbers with all binary strings of the shortest possible length and concatenate these codes. Construct an auxiliary bit sequence of the same length as the main bitstream denoting by '1' the starting positions of codewords. Then a given codeword can be extracted in constant time via known 'select' techniques applied to an auxiliary bit sequence. This method is discussed in \cite{fn1} and called \textit{Simple Dense Coding} (SDC). Its drawback is obvious: although the main bitstream can be quite succinct, the auxiliary bitstream doubles the space which is more than significant.
    \item In a dense sampling scheme \cite{fv} sequence elements are also encoded with all possible binary strings, which constitutes non-uniquely decodable but very dense code. To access the elements two types of pointers are stored: absolute pointers to every $O(\log n)$-th element and relative pointers to the rest. If the integers distribution is not too positively skewed, this makes the constant time direct access possible at the cost of less extra space to the main bitstream, if to compare with the previous approach.
    \item Split binary representations of integers into chunks of a fixed length. Construct the first bitstream from all first chunks, the second bitstream from all second chunks etc. Then the $i$-th element of any bitstream can be accessed on the RAM model in constant time as an array element. However, since the number of chunks in an integer's bit representation is variable, an extra bitmap $B_j$ is used together with the $j$-th bitstream. $B_j[i]=1$ iff the $j$-th chunk of some integer is stored in the $i$-th element of the $j$-th bitstream, and this integer has the $(j+1)$-th chunk. Thus, to reconstruct the $i$-th integer, we get its first chunk from the first bitstream directly and then check the $B_1[i]$. If it is 0, we've got the result, otherwise, we compute the $rank(B_1,i)$ to get the position of the integer's second chunk in the second bitstream and so on. This technique is investigated in \cite{davc} and called the \textit{Directly Addressable variable-length Codes} (DAC). Also it is known as \textit{Reordered Vbyte} since it generalizes the Vbyte code idea \cite{vbyteOrig}. Extraction of a random codeword requires at most $\lceil\frac{\log S}{b}\rceil$ rank operations, where $S$ is the largest integer in the sequence and $b$ is the chunk size. The space overhead for all rank data structures is $O(\frac{n\log\log n}{\log n})$.
    \item Use the variable-length codes with delimiters, for example, Fibonacci codes \cite{AF}, as proposed in another schema from \cite{fn1}. The classical constant-time and $o(n)$-space 'select' algorithm, proposed by Clark \cite{clark}, can be extended to this case.
\end{enumerate}

Our method of fast direct access to elements of a compressed integer sequence is based on the use of variable-length Reverse Multi-delimiter (RMD) Codes introduced in \cite{AZDCC}. These codes are self-delimited, which allows us to avoid using an auxiliary bit-vector indicating codeword boundaries. As well as for Fibonacci codes, it is easy to extend the classical Jacobson's \cite{jacobson} and Clark's \cite{clark} results regarding constant time rank and select on bit-vectors with $o(n)$ extra space to the case of RMD-codes. However, in this case the select operation requires $\frac{3n}{\lceil\lg\lg n\rceil}+O(n^{\frac{1}{2}\lg n\lg\lg n})$ bits of extra space. On the real-world data, for bit sequences of length up to 4Gb, this value exceeds $60\%$ of the code itself, which we consider too big. To reduce space overhead, we combined approaches 1 and 5. Namely, we store the absolute position of each 'Level 1' block of RMD-codewords, then use a kind of linear approximation to get the position of a smaller 'Level 2' block (similar idea has been implemented in \cite{la} and \cite{kulekci}) and search the codeword inside this block sequentially. Properties of RMD-codes allow us to perform this search and decoding significantly faster and use smaller space than for Elias codes.

The variable-length data compression RMD-codes and their properties are discussed in Section \ref{rmd}. The main method of integer retrieval is presented in Section \ref{Algo}. Its space complexity is estimated in Section \ref{space}. In Section \ref{exp} we discuss experiments in compression and extracting elements from an integer sequence generated in the process of English text compression. As shown in \cite{AZDCC}, RMD-codes outperform Fibonacci codes both in natural language text compression ratio and in decoding speed. That is why we did not include in experiments the Fibonacci-based approach \cite{fn1}. Also, in practical English text compression, DAC outperforms the dense sampling both by space and time, as shown in \cite{davc}. Thus, we also exclude the dense sampling scheme from our experimental set, where remain the SDC encoding, DAC, our new method, and, as the base of comparison, naive approach 1 implemented with the use of Elias delta code. 

\section{Reverse Multi-Delimiter Codes}
\label{rmd}
Let $\mathcal{M}=\{m_{1},\ldots,m_{t}\}$  be a set of positive integers, given in ascending order.

\begin{defin}
The reverse multi-delimiter code $R_{m_1,\ldots,m_t}$ consists of all the words of the form  $01^{m_{i}}, i=1,\ldots,t$ and all other words that meet the following requirements:

\begin{enumerate}[itemindent=1cm,leftmargin=0cm,itemsep=0cm,parsep=0.1cm]
\item[(i)] for any $m_{i}\in\mathcal{M}$ a word does not end with the sequence $01^{m_{i}}$;
\item[(ii)] for any $m_{i}\in\mathcal{M}$ a word can contain the sequence $01^{m_{i}}0$ only as a prefix;
\item[(iii)] a word starts with the prefix $01^{m_{i}}0$ for some $m_{i}\in\mathcal{M}$.
\end{enumerate}
\end{defin}

The given definition implies that code delimiters in $R_{m_1,\ldots,m_t}$ are sequences of the form $01^{m_{i}}0$. However, the code also contains shorter words of the form $01^{m_{i}}$ that form a delimiter together with the first zero of the next codeword.

For RMD-codes, there exists a monotonic encoding mapping from the set of natural numbers to the set of codewords. First, this was announced in \cite{AZDCC}. For the sake of completeness, we describe this useful construction that is a base for very efficient decompression procedures.
Let $K=\{k_1,\ldots,k_q\}$ be the ascending sequence of all integers in the range $[0,m_t+1]$ that don't belong to $\mathcal{M}$. For instance, $K=\{0,1,3,6\}$ for the code $R_{2,4,5}$. Note that each codeword of $R_{m_1,\ldots,m_t}$ has the following structure: it consists of a prefix $01^{m_{i}}$, for some $m_i\in\mathcal{M}$, which can be followed by some groups of bits having the form $01^s$, where $s\in K$ or $s>k_q$. The inverse statement is also correct: any bit sequence of the described structure forms a codeword.

Therefore, in the code $R_{m_1,\ldots,m_t}$ any codeword of the length $L$ can be constructed in one (and only one) of the following ways:

\begin{enumerate}
\item[(i)] it is composed of a codeword of the length $L-s-1$ followed by the sequence $01^s, s\in K$;

\item[(ii)] it is composed of a codeword of the length $L-1$ with a suffix $01^r, r>0$, appended with the single '1' bit (note that neither $r$ nor $r+1$ belong to $\mathcal{M}$);

\item[(iii)] it is a word of the form $01^{m_{i}}, i=1,\ldots,t$.
\end{enumerate}

Based on these facts, we can formulate the principle of constructing the ordered set of codewords of the length $L$ when the sets of shorter codewords are already built.

\begin{enumerate}
\item[(i)] Iterating $i$ from 1 to $q$, replicate the sets of codewords of lengths $L-k_i-1$ and append sequences of the form $01^{k_i}$, $k_i\in K$, to them.

\item[(ii)] Replicate the set of words of the length $L-1$ with a suffix $01^r$, where $r>0$ and $r+1\notin\mathcal{M}$, and append the single '1' bit to all elements of this set.

\item[(iii)] If $L=m_i+1$, $i\in \{1,\ldots,t\}$, append the word $01^{m_i}$ to the codeword set.
\end{enumerate}

\begin{figure}[h!]
\resizebox{\textwidth}{!}{
  \includegraphics[trim=27cm 1cm 1cm 1cm,clip]{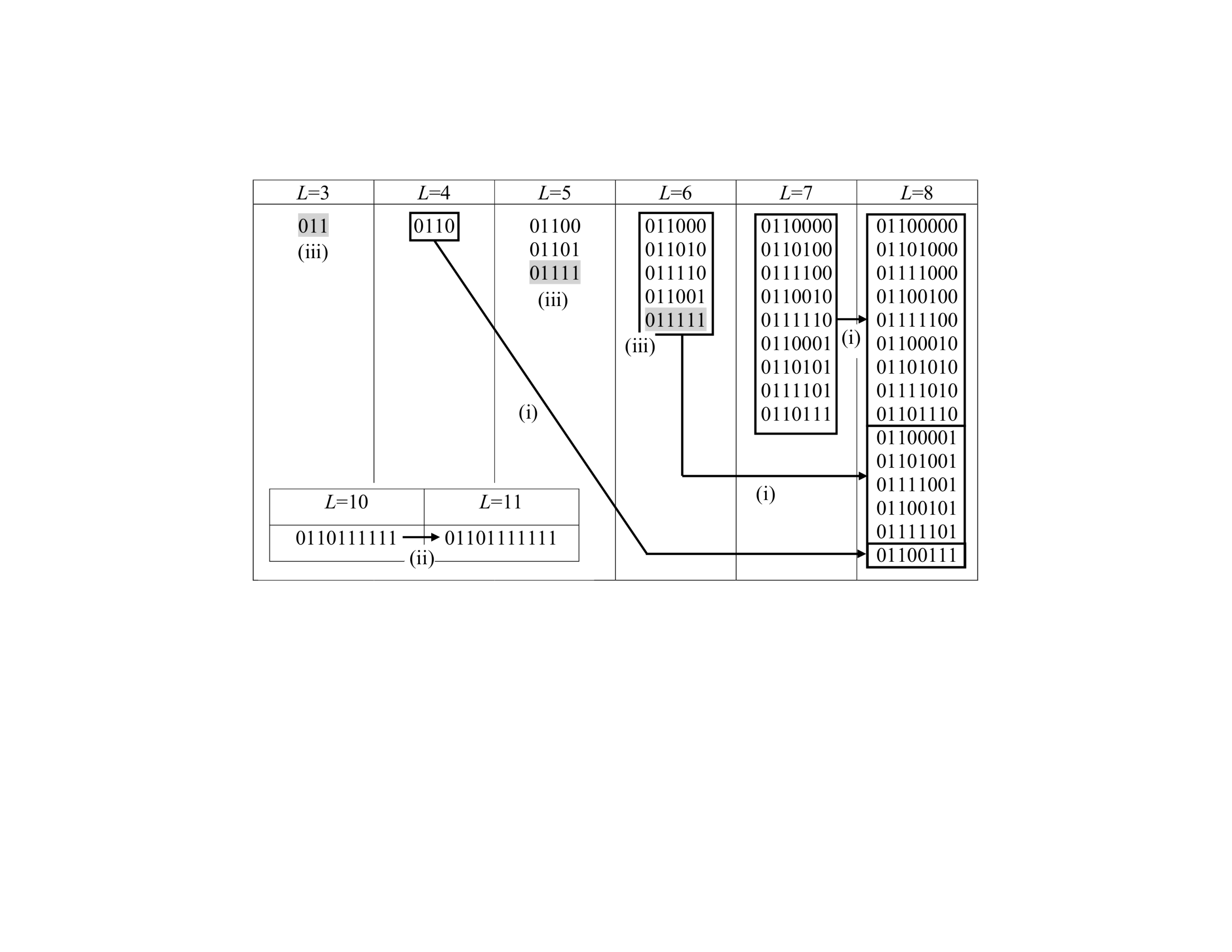}
}
\caption{$R_{2,4,5}$ codewords construction}
\label{fig:1}
\end{figure}

If we step by step apply this approach to form the sets of codewords of lengths $m_1+1,m_1+2,\ldots$, we get the set of all $R_{m_1,\ldots,m_t}$-codewords ordered by length ascending. For example, the set of $R_{2,4,5}$-codewords of length 8 or less is given in Fig. \ref{fig:1}. Also, it demonstrates how the words of length 8 are constructed from the words of lengths 7, 6, and 4 by appending bit sequences 0, 01, and 0111 respectively (rule 1). Three codewords highlighted in grey are constructed by applying rule 3. The shortest codeword that is constructed by applying rule 2 has length 11; its construction is shown in the left bottom part of the figure.

Any reverse multi-delimiter code $R_{m_1,\ldots,m_t}$ contains the same number of codewords of a given length as the ``direct'' multi-delimiter code $D_{m_1,\ldots,m_t}$ discussed in \cite{AZ}. Thus, reverse MD-codes possess all properties of MD-codes such as completeness and universality as well as their asymptotic densities. For MD-codes, these properties were proven in \cite{AZ}. Also, we refer to \cite{AZ} for the analysis of asymptotic densities and quantities of short codewords in multi-delimiter codes.

In the sequel, we concentrate on the “infinite” versions of RMD-codes, notably $R_{2-\infty}$ and $R_{2,4-\infty}$, as they demonstrate the best compression ratio. They use all delimiters containing $2,3,\ldots$ or $2,4,5,\ldots$ ones respectively. However, in practice it is enough to limit lengths of delimiters by some relatively large number, defined by the maximal codeword length for specific application.

As mentioned above, we need to decode the RMD-encoded numbers to retrieve them from a compressed sequence. An RMD-code can be considered as a regular language and thus recognized by the finite automaton. The decoding automata for codes $R_{2-\infty}$, $R_{3-\infty}$, and $R_{2,4-\infty}$ are given and discussed in \cite{vz}. However, they process a text bit-by-bit, which is quite slow. The main idea of a fast decoding algorithm is a ``quantification'' of a decoding automaton so that it reads bytes of a code and produces the corresponding output numbers. Such an algorithm has been proposed in \cite{AZDCC}. It analyzes how many codewords are decoded at each iteration of a decoding loop and produces the corresponding outputs. Unfortunately, the ’if’ statements used for that purpose are unpredictable (i.e. can be either ’true’ or ’false’ with high probability) and this slows down the algorithm on modern processors. However, we can avoid performing conditional statements by exploiting an idea similar to the fast decoding of a binary-encoded ternary code given in \cite{ternary}. We use this approach in Algorithm \ref{a2} described in the next Section, which is a part of general integer retrieving Algorithm \ref{a1}.

\section{Integer Retrieving Technique}
\label{Algo}

Below we describe Algorithm \ref{a1} calculating the value of the element with a given index in the integer sequence encoded with RMD-codes. Hereinafter we consider codes $R_{2,x}$ having the shortest delimiter 0110; they are the best representatives of an RMD family in natural language text compression. Although an RMD-encoded sequence is a bitstream, to make the method fast, we operate on a byte level getting all required bit-level data from lookup tables. The algorithm idea and notations are the following.

\begin{itemize}[itemsep=0cm,parsep=0.2cm]
    \item Split the encoded bitstream into level 1 blocks containing $L_1$ codewords each. Store the number of the first byte of each block in the array $L1byte$.
    \item Split each L1-block into level 2 blocks containing $L_2$ codewords each. Let $L2Length[i]$ be the average length in bytes of an L2-block in the $i$-th L1-block. Also, store the array $\Delta_b[i][j]=b_{ij}-\lfloor j\cdot L2Length[i]\rfloor$, where $b_{ij}$ is the position of the first byte of the $j$-th level 2 block relative to $i$-th level 1 block. Then the number of the leftmost byte of the L2-block we can calculate by the formula $L1byte[i]+j\cdot L2Length[i]+\Delta_b[i][j]$. As shown in \cite{AZDCC}, no more than 3 codewords of an RMD-code $R_{2,x}$ can start in one byte. Therefore, we need also the 2-bit value $\Delta_c[i][j]$ indicating which codeword inside the byte is the first codeword of the $j$-th level 2 block.
    \item When we know exactly where the L2-block starts, seek the element inside the block processing it byte-by-byte. It can be done from the beginning of the block in the left-to-right direction or from the beginning of the next block right-to-left, depending on which way is shorter.
    \item When we found the leftmost byte of a required codeword, decode it using a fast byte-aligned decoding technique, e.g. as discussed in \cite{ternary}.
\end{itemize}

Note that in general the $j$-th L2-block position can be approximated by the formula $kj+b$, where parameters $k$ and $b$ are calculated with the ordinary least squares technique. However, we intentionally fix $b$ as $L1byte[i]$ since experiments show that this approach is a bit less space-consuming.

\begin{algorithm}[H]
\label{a1}
\SetKwInOut{Input}{input}\SetKwInOut{Output}{output}
\caption{Decoding an element of the RMD-encoded sequence}
\Input{The index $t$ of the element.}
\Output{The value of the $t$-th element, $out$.}
$n_1\lra t\,\textrm{div}\,L_1$\tcp*[f]{Number of the L1-block}\\
$e_1\lra t\,\textrm{mod}\,L_1$\tcp*[f]{Number of the element inside the L1-block}\\
$n_2\lra e_1\,\textrm{div}\,L_2$\tcp*[f]{Number of the L2-block inside the L1-block}\\

\tcp{Number of the codeword inside the byte-aligned L2-block}
$e_2\lra e_1\,\textrm{mod}\,L_2+\Delta_c[n_1][n_2]$\\
\tcp{Find the byte where the codeword starts}
\eIf{$e_2<L_2/2$}
{
\tcp{Number of the byte where the L2-block starts}
$i\lra L1byte[n_1]+L2Length[n_1]*n_2+\Delta_b[n_1][n_2]$\\
$e\lra0$}
{
$n_2\lra n_2+1$\tcp*[f]{Search from the next L2-block right to left}\\
$i\lra L1byte[n_1]+L2Length[n_1]*n_2+\Delta_b[n_1][n_2]-1$\\
$e\lra L_2+\Delta_c[n_1][n_2-1]-\Delta_c[n_1][n_2]$\\
\While{$e\geq e_2$}{
$e\lra e-Words(i)$\\
$i\lra i-1$\\}
}
\While{$e<e_2$}{
$e\lra e+Words(i)$\\
$i\lra i+1$\\}
\tcp{Starting from the $(i-1)$-th byte of a code, skip $e-e_2$ codewords, and decode the next codeword }
$out\lra Decode\_number(i-1,e-e_2)$\\
\end{algorithm}

Let us explain how Alg. \ref{a1} works. Given the index $t$ of a required element, in lines 1 and 2 we get the number of the containing L1-block, $n_1$, and the relative number of the element inside this L1-block, $e_1$. Consider the L2-block containing the required codeword. Its number $n_2$ relative to the containing L1-block is calculated in line 3 of Alg. \ref{a1}. 

We search a codeword inside the L2-block byte-by-byte. However, an L2-block may start not from the first codeword in the byte, but from the second or third. Then we extend the discussed L2-block to the left by including all full codewords from its first byte and call this extended block a \textit{byte-aligned L2-block}. The difference between the codeword positions in the byte-aligned and original L2-blocks we store in the array $\Delta_c$, and get the number of the required codeword inside the byte-aligned L2-block in line 4 of Alg. \ref{a1} denoting it by $e_2$.

\begin{figure}
\centering
\includegraphics[scale=0.8,clip=true]{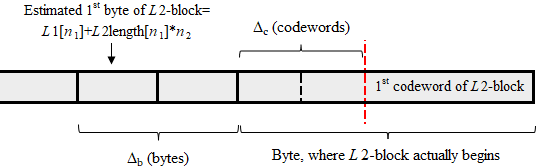}
\caption{Calculating the position of an L2-block} \label{fig0}
\end{figure}

In line 5 we analyze if the codeword is in the left half of the L2-block. If so, using a kind of linear approximation, in line 6 we get the number of the byte where this L2-block starts. Then in lines 6-7 and 17-20 the number $i$ of the first byte of a required codeword is calculated by sequential processing bytes of L2-block from left to right. The function $Words(i)$ returns the number of codewords starting in the $i$-th byte of a code. It is summed up in the variable $e$ until it becomes no less than the required value $e_2$. The correspondence between the estimated and actual beginning of an L2-block, as well as values from $\Delta_b$ and $\Delta_c$ arrays, is shown in Fig. \ref{fig0}. 

If the required codeword is in the right half of the L2-block (lines 9-15), the right-to-left search from the beginning of the next L2-block would be faster. In this case, we increment the number of the L2-block (line 9), get the number of the byte before its beginning (line 10) and the number of codewords the right-to-left search to be started from (line 11). After the search finishes, the value $e$ may become too small and a few iterations of the left-to-right search may be needed (lines 17-20).

In both cases, after line 20, the index $i-1$ points to the byte where the required codeword starts, and it is the $(e-e_2)$-th codeword in this byte if we count from 0 and from the right edge of the byte. Thus, we skip $e-e_2$ codewords from the right edge of the byte and return the result of decoding the next codeword to the left of them (line 21). This is done in the function $Decode\_number$, which is described in Alg. \ref{a2}. Its idea resembles the fast decoding method given in \cite{ternary}.

\begin{algorithm}[H]
\label{a2}
\SetKwInOut{Input}{input}\SetKwInOut{Output}{output}
\caption{Function $Decode\_number(i,s)$ - decoding the $(s+1)$-th \\codeword starting from the right edge of the $i$-th byte of a code (little-endian machine)}
\Input{$i$ - the number of the byte of a code; $s\leq2$ - the number of codewords to be skipped. }
\Output{The decoded number, $out$.}
$val64\lra Code[i...i+7]$\tcp*[f]{Read 8 bytes of a code}\\
$val64\lra Align(val64,s)$\tcp*[f]{Mask out the rightmost $s$ codewords and align the next codeword to the right edge of a 64-bit word}\\
$ptr\lra0$; $out\lra0$\\
$chunk\_mask\lra2^{chunk\_size}-1$\\
\Repeat{$N[v]=0$}{
$v\lra ptr<<chunk\_size+val64\&chunk\_mask$\\
$ptr\lra Pointers[v]$\\
$out\lra out+Out[v]$\\
$val64\lra val64>>chunk\_size$
}
\end{algorithm}

Assume a real-world text codeword is never longer than 57 bits. In lines 1 and 2, we load 8 bytes of a code containing the whole codeword to be decoded into the variable $val64$ and shift this codeword to the right edge of a 64-bit machine word. Then we split the bit representation of $val64$ into chunks and process it chunk-by-chunk  accumulating the resultant value in the variable $out$. Alg. \ref{a2} demonstrates this process on a little-endian machine, where bytes of a value are loaded from memory to a processor register in the reverse order, and thus the bits inside bytes of a code should also be put in the reverse order.

The chunks of an RMD-encoded bitstream are recognized by quantified finite automatons, as described in \cite{AZDCC}. The result of the chunk decoding depends on the content of the chunk, the number of the chunk, and the state of the decoding automaton at the beginning of chunk processing. The latter 2 parameters are stored in the variable $ptr$ used in lines 6 and 7. In line 6, we shift its value by the chunk bitsize to the left and add the chunk content to it. This way we obtain the value $v$ containing the full information to decode the current chunk (line 6). Then, in line 7 we get the value $ptr$ for the next chunk and increment the current result by the value $Out[v]$ in line 8. At last, in line 9 the value $val64$ is shifted by the chunk size to the right to process the next chunk. The process repeats until the flag $N[v]$ signals that we met a delimiter and the codeword has been decoded.

\begin{example}
Assume $L_1=2^{10}=1024$ and $L_2={2^5}=32$ and retrieve the 1060-th element from the compressed integer sequence encoded by $R_{2,4-\infty}$. At first, we get $n_1=1060\,\textrm{div}\,1024=1$ (number of L1-block), $e_1= 1060\,\textrm{mod}\,1024=36$ (number of the byte inside the L1-block), and $n_2=36\,\textrm{div}\,32=1$ (number of L2-block). Then, assume the first L1-block occupies 1200 full bytes of an RMD-bitstream, i.e. $L1byte[1]=1200$, and the average length of an L2-block inside the second L1-block is 40 bytes, i.e. $L2Length[1]=40$. However, the actual byte length of the first L2-block inside the second L1-block can be different, say 39. Then $\Delta_b[1][1]=39-40=-1$. For example, this block can occupy some rightmost bits of the 1200-th byte, full bytes $1201-1238$, and 5 leftmost bits of the byte 1239, as shown in Fig. \ref{fig:ex}. 

Now, assume the binary representation of the 1239-th byte is $00011011$. The leftmost 2 bits represent the ending of the 1055-th codeword and together with the 1056-th codeword $011$ belong to the first L2-block inside the second L1-block. Then the last 3 bits $011$ are the starting bits of the next L2-block, which is of our interest. Since this L2-block starts from the second codeword in the byte 1239, $\Delta_c[1][1]=1$, i.e. we should skip one full codeword in the byte 1239 to get to the beginning of the L2-block.
Then $e_2=e_1\,\textrm{mod}\,L_2+\Delta_c[n_1][n_2]=(36\,\textrm{mod}\,32)+1=5$ is the number of the target codeword if we start counting from the first full codeword in the byte 1239. 

Since $e_2<L_2/2$, we execute lines 6 and 7 of Algorithm \ref{a1}: $i\lra L1byte[n_1]+L2Length[n_1]*n_2+\Delta_b[n_1][n_2]=1200+40*1-1=1239$, $e=0$. Then we execute iterations of the loop in lines $17-19$ assuming the bitstream is shown in Fig. \ref{fig:ex}, where even bytes are highlighted with grey. 
\begin{enumerate}
\item $Words(1239)=2$, $e=2$, $i=1240$;
\item $Words(1240)=2$, $e=4$, $i=1241$;
\item $Words(1241)=0$, $e=4$, $i=1242$;
\item $Words(1242)=2$, $e=6$, $i=1243$;
\end{enumerate}
At last, we call the function $Decode\_number(i-1,e-e_2)=Decode\_number(1242,1)$. It skips 1 codeword (1061-th) from the right edge of byte 1242, aligns the 1060-th codeword $01101$ to the right edge of a 64-bit machine word, and returns its value, i.e. 3 (see Fig. \ref{fig:1}).

\begin{figure}[h!]
\resizebox{\textwidth}{!}{
  \includegraphics[width=\textwidth]{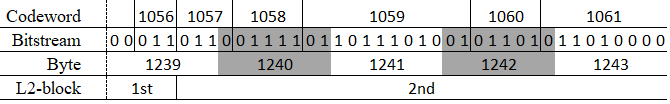}
}
\caption{Fragment of an $R_{2,4-\infty}$-bitstream}
\label{fig:ex}
\end{figure}
\end{example}

\section{Space complexity}
\label{space}
Now, let us estimate the space required by Algorithms \ref{a1} and \ref{a2}, apart from the size of an RMD-encoded file itself. Assume the encoded sequence fits into 4GB, and $n$ is the number of elements in it. Then it is enough 4 bytes to store an $L1byte$ array element, or $4n/L_1$ bytes for the whole $L1byte$ array. If we reserve 4 bytes to store the linear approximation ratio $L2Length[n_1]$, the array $L2Length$ will occupy the same space. As mentioned above, no more than 3 codewords of $R_{2-x}$ code can start in one byte. Therefore, it is enough 2 bits for an element of the array $\Delta_c$, or $n/4L_2$ bytes for the whole array. 

The function $Word(i)$ uses the lookup table consisting of the number of codewords starting in the byte $code[i]$. Analyzing the byte itself, it is not possible to determine how many codewords start in it. For example, if the byte ends with a 0 bit, it can be either the first bit of the next codeword or a continuation of the current one. However, to answer this question for the code $R_{2-\infty}$ we need to analyze only 2 bits following the current byte, and 4 bits for the code $R_{2,4-\infty}$. Namely, the sequences $0|11$ in $R_{2-\infty}$, and $0|110$ or $0|1111$ in $R_{2,4-\infty}$ begin the new codeword, while all other bit combinations after the ending 0 mean that the current codeword continues. For ending 1, we need to analyze even fewer extra bits. Thus, the index of the mentioned lookup table can be a 12-bit integer, and the table consists of 4096 elements. To decrease the number of bit-level operations, we reserve 1 byte for each element (the number between 0 and 3), and 4096 bytes will be enough to store the whole table.

To estimate the space complexity of Alg. \ref{a2}, we should calculate the maximal value of the variable $v$. If a codeword consists of not more than $max\_len$ bits, it contains not more than $c=\lceil \frac{max\_len}{chunk\_size}\rceil$ chunks. As mentioned above, the result of the decoding depends on the number of the chunk, its content, and the state of the decoding automaton. Thus, $v\leq c\cdot n\_states\cdot2^{chunk\_size}$, where $n\_states$ is the number of states in the decoding automaton (3 for $R_{2-\infty}$ and 5 for $R_{2,4-\infty}$ \cite{vz}). Assuming the realistic codeword length does not exceed 40-45 bits, and considering the value $chunk\_size=7$, which gives the lowest decoding time in experiments, we get $4000-5000$ as an upper bound for $v$. Each element of the arrays $Pointers$ and $N$ takes 1 byte, while $Out[v]$ requires 4 bytes. Therefore, the total size of the lookup tables for Algorithm \ref{a2} does not exceed $25-30$KB. 

The array $\Delta_b$ occupies the biggest space. These delta values can vary in different ranges for different L1-blocks. That is why we allocate the different number of bits for elements of different $\Delta_b[i]$ subarrays and store these bitlengths in the special array $Bit\_ranges$. All $\Delta_b$ values for an $L1$-block we store as a bitstream, also keeping a pointer to it in the array $\Delta ptr$. It is enough 2 bytes for a $Bit\_ranges$ array element and 4 bytes for a pointer. Of course, this approach involves a number of extra bit operations. Nonetheless, it allows us to save $40-50\%$ space occupied by data structures needed for direct access and does not affect the overall time much because the biggest time consumption is accounted for the loops in lines $12-20$ of Algorithm \ref{a1}. Also, using smaller data structures accelerates an algorithm thanks to fewer cache mismatches.

In total, we need $25-30$KB of memory for Alg. \ref{a2}, $14n/L_1$ bytes for level 1 structures, $n/4L_2$ bytes for the array $\Delta_c$, and a variable space for the array $\Delta_b$. As shown in experiments described in the next section, the optimal value for $L_1$ can be between $2^{14}$ and $2^{17}$, while for $L_2$ it is between $2^6$ and $2^8$. This makes the space for $L1$-structures and $\Delta_c$ almost insignificant, about tenths of 1 percent of a code itself, while $\Delta_b$ occupies about $1-3\%$ of the code size.

\section{Experiments}
\label{exp}
We tested our solution on integer sequences obtained by applying two known natural language compression schemes to 200MB English text from Pizza\&Chili corpus.

\begin{itemize}
\item In the first scheme, words of the text are considered as alphabet symbols. In the dictionary, they are arranged in the order of descending frequencies. Then we replace words in the text with their indices in the dictionary. The text consists of 37,003,242 words and has the entropy H0 52,805 KB.
\item The second scheme was proposed by Ferragina and Venturini \cite{fv} to compress a sequence of $n$ characters to its high-order entropy so that a $O(\log n)$-bit substring can be decoded in constant time. The text is split into blocks of $\frac{1}{2}\log n$ bits, which are sorted by frequency and encoded as in the first scheme. In our test $n=209,715,202$, which implies $\frac{1}{2}\log n\approx14$. In order to reduce the volume of bit-level operations, we rounded the block size to 2 bytes. Since alphabet is constructed of pairs of characters, the compressed text size should be compared with entropy H1, which is 106,754 KB.
\end{itemize}

We measured the element extraction time as well as the space occupied both by the encoded text and the auxiliary structures. The time was averaged over 100 mln. extractions of a random integer sequence element. To reduce the number of divisions in Algorithm \ref{a1}, we chose the size both of L1 and L2-blocks as powers of 2: $L_1=2^{l_1}$ and $L_2=2^{l_2}$. The optimal chunk size in Algorithm \ref{a2} was determined experimentally and equals 7 bits for all tests.

\begin{figure}[H]
\includegraphics[width=0.8\textwidth]{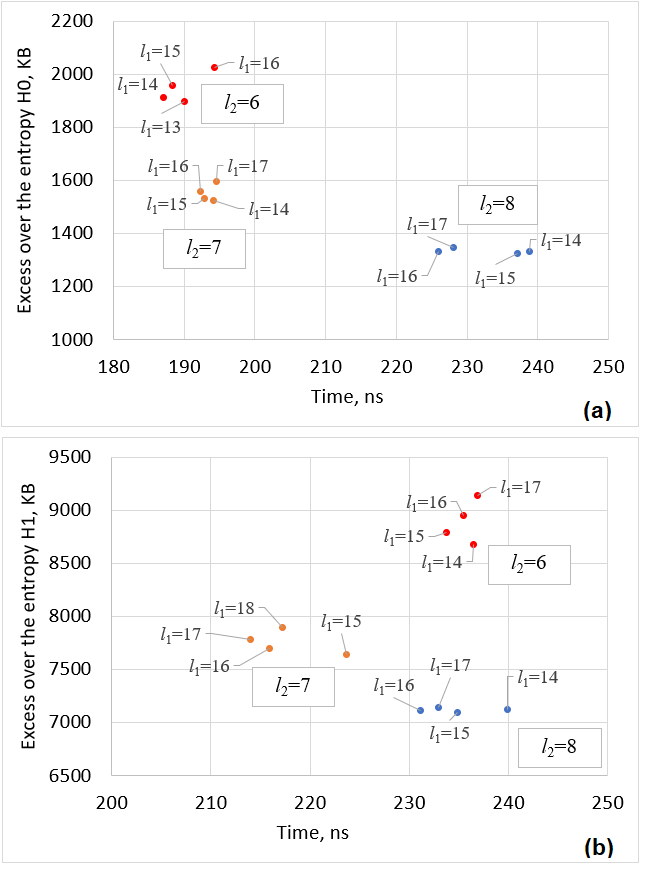}
\caption{Element extracting from RMD-bitstream: (a) word-based alphabet, (b) character-based alphabet} \label{fig1}
\end{figure}

Parameters $l_1$ and $l_2$ constitute a space/time trade-off shown in Fig. \ref{fig1}. Parameter $l_2$ has more impact because it defines the average number of iterations of the loops in Algorithm 1 as well as the size of arrays $\Delta_c$ and $\Delta_b$. As mentioned above, the most prominent values of $l_2$ for our data are between 6 and 8. When $l_2$ decreases, arrays $\Delta_c$ and $\Delta_b$ become bigger but loops have fewer iterations. This speeds up the algorithm by the cost of space until arrays become too big to fit into the L2 or L3 cache, which causes many cache mismatches. The latter situation is illustrated in Fig. \ref{fig1}b, where the element extraction for $l_2=6$ is both longer and requires more space than for $l_2=7$. 

Two competitive solutions discussed in the Introduction were tested for the comparison: the Directly Addressable variable-length Codes (DAC) \cite{davc} and a Simple Dense Coding (SDC) \cite{fn1}. The DAC relies on a 'black-box' rank operation for a bit-sequence, while random access via SDC structure requires a select. The comparison of the best recent approaches to computing rank and select for binary sequences is given in \cite{kulekci} as well as in \cite{pibiri}. In both sources, the two fastest methods to compute rank appear to be Rank9-V1 \cite{vigna} and its variation, the so-called IL (interleaving) \cite{gog}, where the original bit-vector is interleaved with rank information. They also require relatively small space overhead (usually Rank9-V1 uses somewhat bigger space than IL). 

As reported in \cite{kulekci}, very fast 'select' algorithms operating at approximately the same speed are provided with SD \cite{sd}, MCL, RSAA \cite{kulekci}, and LA \cite{la} structures. However, the space complexity highly depends on the percentage of ones in a bit-vector. In the first scheme of our test set it is about 13\%, and about 19\% in the second scheme. For bit-vectors with low percentage of ones, SD and LA select-structures occupy more attractive positions on the space/time plane than the other two mentioned solutions. To implement them, the simple dense code of the integer sequence we store as-is, while the auxiliary binary sequence needed for constant time random access is given in the form of a compressed LA- or SD-vector.

Also, we tested the naive method mentioned as "approach 1" in Introduction. The integer sequences were encoded with the Elias $\delta$-code and the position of each $s$-th codeword was sampled. To retrieve some element, we perform the sequential search from the sampled position. 

\begin{figure}[H]
\includegraphics[width=0.8\textwidth]{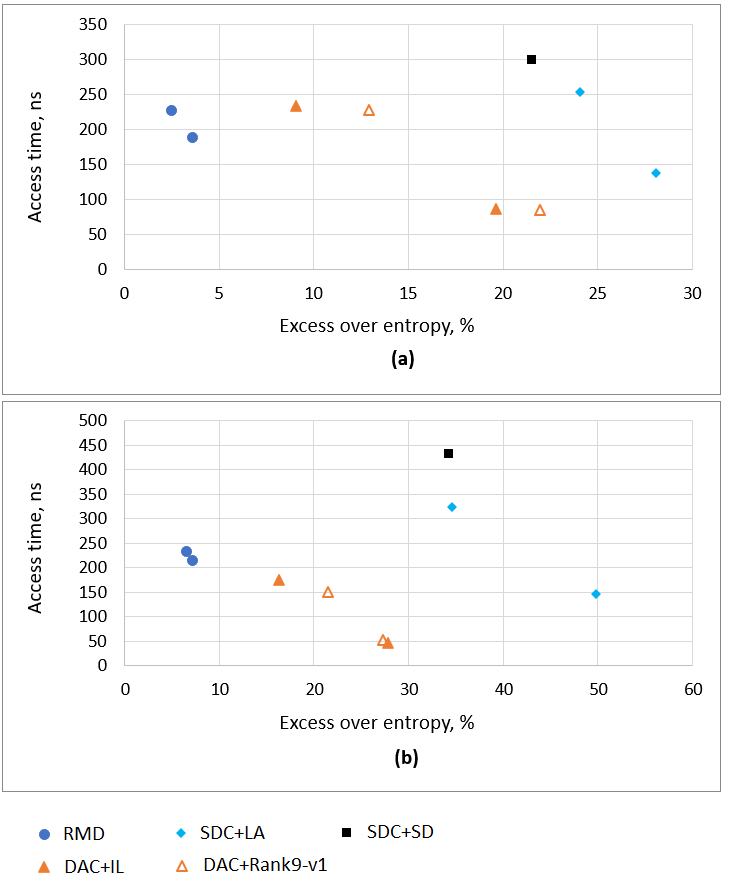}
\caption{Experiments with different approaches: (a) word-based alphabet, (b) character-based alphabet} \label{fig5}
\end{figure}

The compressed file sizes together with auxiliary data structures as well as average integer extraction times are shown in Table \ref{tab1} and in Fig. \ref{fig5} (the naive approach is not shown in the Figure as it goes beyond the scale). The excess over the $H_0$ entropy for the word alphabet and over the $H_1$ entropy for the character-based one is shown in percentage. All data is stored in RAM. To build our and other solutions we used the g++ compiler v9.4.0 with -O3 optimization flag. We got IL and SD implementations from the SDSL library \cite{sdsl}, and LA implementation from \cite{lagit}. Tests have been run on a computer with an AMD Athlon 3000G processor, 32 KB of L1 cache, 512 KB of L2 cache, 4 MB of L3 cache, 16 GB RAM, and OS Ubuntu 20.04 LTS. The source code can be downloaded from \cite{gitra}.

We tested methods with different parameters representing different points in the space/time trade-off. For the first scheme, the code $R_{2,4-\infty}$ gives the best compression ratio, while for the second scheme it is $R_{2-\infty}$. In each case, we show two pairs of parameters $l_1,l_2$ giving the best time and the best space, as well as space or time optimal LA-parameters for the SDC+LA scheme. The DAC code is parameterized by the bit size of the chunk $b$. The Elias $\delta$-code performance depends on the sampling interval $s$.

\begin{table*}[!t]
\caption{Experiments on integer compression and extraction}
\label{tab1}
\centering
Integer sequence generated from the word-based alphabet\\
\begin{tabular}
{|c|c|c|c|}
\hline
Algorithm&Parameters&Size, KB&Time, ns\\ \hline
\multirow{2}{*}{Elias $\delta$}&$s=4$&107,835\,(104.9\%)&127\\
&$s=512$&71,121\,(34.69\%)&1999\\ \hline
\multirow{2}{*}{RMD, $R_{2,4-\infty}$}&$l_1=14,\,l_2=6$&54,719\,(3.62\%)&187\\ 
&$l_1=16,\,l_2=8$&54,138\,(2.52\%)&226\\ \hline
\multirow{4}{*}{DAC}&$b=8$, IL&\,63,163\,(19.61\%)\;&86\\
&$b=8$, Rank9-v1&64,387\,(21.93\%)&85\\
&$b=4$, IL&57,595\,(9.07\%)&233\\ 
&$b=4$, Rank9-v1&59,628\,(12.92\%)&228\\ \hline
\multirow{2}{*}{SDC+LA}&$bpc=7$&67,634\,(28.08\%)&137\\ 
&$v$-$opt$&65,500\,(24.04\%)&253\\ \hline
SDC+SD&&64,164\,(21.51\%)&300\\ \hline
\end{tabular}
\vspace{2pt}
\\Integer sequence generated from the character-based alphabet\\
\vspace{2pt}
\begin{tabular}{|c|c|c|c|}
\hline
\multirow{2}{*}{Elias $\delta$}&$s=4$&252,732\,(136.7\%)&135\\
&$s=512$&148,694\,(39.3\%)&1932\\ \hline
\multirow{2}{*}{RMD, $R_{2-\infty}\,\,\;$}&$l_1=17,\,l_2=7$&114,538\,(7.29\%)&$\quad\,$214$\quad\,$\\ 
&$l_1=16,\,l_2=8$&113,863\,(6.65\%)&231\\ \hline
\multirow{4}{*}{DAC}&$b=8$, IL&136,444\,(27.81\%)&47\\
&$b=8$, Rank9-v1&135,935\,(27.33\%)&53\\
&$b=4$, IL&124,133\,(16.28\%)&175\\
&$b=4$, Rank9-v1&129,734\,(21.5\%)&150\\\hline
\multirow{2}{*}{SDC+LA}&$bpc=7$&159,914\,(49.8\%)&146\\ 
&$v$-$opt$&143,696\,(34.61\%)&323\\ \hline
SDC+SD&&143,301\,(34.23\%)&433\\ \hline
\end{tabular}
\end{table*}

As seen, the Elias codes are obviously space inefficient. Even the code itself exceeds the entropy H0 by 34\% for word-based alphabet and by 38\% for character-based. However, the extraction time can be quite low if the sampling rate is high. In fact, this way we approach the uncompressed integer sequence.

Our data structure based on RMD-codes is significantly more compact than all competitive solutions both for word-based and character-based alphabets. Also, our method of random access is faster than SDC in combination with the space-optimal LA or SD on both alphabets. However, SDC+LA scheme may become faster at the cost of extra space ($bpc=7$).

We tested DAC with different bit sizes of a chunk: $b=4$ or $b=8$. This parameter represents a space/time trade-off: operating whole bytes ($b=8$) is much faster but requires much more space than for $b=4$. The value $b=2$ appears to be not efficient and is not shown in the table. 

Our method is better than DAC-4 both in space and time on the word-based alphabet: $3-10\%$ shorter and $1-25\%$ faster, depending on RMD parameters and variants of DAC. Enlarging the chunk size to 8 bits makes DAC $2-2.7$ times faster than RMD by the cost of space (it becomes $15-19\%$ larger).

On the character-based alphabet, the encoded text becomes larger, and the size of the arrays $L1Byte$, $L2Length$, $\Delta_c$, and $\Delta_b$ also grows causing more cache mismatches and slowing down our algorithm. At the same time, encoded integers are smaller requiring less number of streams in DAC. As a result, on the character-based alphabet, our method becomes slower even than DAC-4. However, its space outperformance increases (the space optimal RMD structure is shorter than DAC by $9-14\%$ for $b=4$ and by $19-20\%$ for $b=8$).

\section{Conclusion}

We presented a fast method of extracting an element of an unordered integer sequence compressed with the use of Reverse Multi-Delimiter codes. By exploiting the recently developed technique of linear approximation of a codeword block position and properties of RMD-codes, we achieved a very good compression ratio by taking experimental integer sequences from a frequency-based compression of the 200MB English text. Together with all data structures required for fast direct access, the size of the compressed file exceeds the zero-order entropy on the word-based alphabet by $2.5-3.5\%$ and the first-order entropy on the character-based alphabet by $6.5-7.5\%$. At the same time, our method provides a decent speed of element extraction, being the fastest among competitive solutions that compress the text with the ratio exceeding the entropy by less than 15\%.

\bibliographystyle{IEEEtran}
\bibliography{IEEEabrv,bib}
\end{document}